\documentclass[a4paper,10pt]{article}
\usepackage{latexsym}
\usepackage{amsmath}
\usepackage{amssymb}
\usepackage{graphicx}
\usepackage{wrapfig}
\usepackage{fancybox}
\usepackage{bm}
\title{Spectral Functions at finite temperature and chemical potential}
\author{S.Sasagawa and H.Tanaka \\\\Department of Physics, Rikkyo University, Tokyo 171-8501, Japan}
\date{}
\begin{document}
\maketitle

\vspace{1em}
\vspace{1em}
\vspace{1em}
\begin{abstract}

There are two formulations at non-zero chemical potential; one is the formulation that a Lagrangian includes a chemical potential, the other is the formulation that a Lagrangian does not include a chemical potential. The existence of two formulations makes a calculation complicated. The results from those formulations are not corresponding directly. This discrepancy exists in the imaginary time formalism and the real time formalism. However, since this is essentially caused by a difference of a frequency, the discrepancy vanishes by modifying the Fourier transform. We show a calculational procedure with a spectral function to understand this.

\end{abstract}

\vspace{1em}
\vspace{1em}
\section{Introduction}

As shown in Ref.\ \cite{rf:weldon2}, there is a cumbersome puzzle at finite temperature and chemical potential. The puzzle is caused by the existence of two formulations in the real time formalism at non-zero chemical potential\cite{rf:weldon2}\cite{rf:kobe}. This raises a problem on a specific calculation. For example, a problem appears on a calculation of a summation at finite temperature and chemical potential. However, if the imaginary time formalism\cite{rf:fetter}\cite{rf:kapsta} is consistently used, this problem does not emerge. It emerges when a method with real time is adopted. In particular, it emerges more directly in the real time formalism\cite{rf:weldon2}\cite{rf:kobe}. For this reason, clarifying details of the puzzle is important to maintain the consistency between imaginary time and real time.

There are two formulations at non-zero chemical potential; one is the formulation (A) that a Lagrangian includes a chemical potential\cite{rf:weldon1}, the other is the formulation (B) that a Lagrangian does not include a chemical potential\cite{rf:kobe}. A problem does not emerge as long as only either is used in the real time formalism. However, when one verifies whether a result agrees with a one from the imaginary time formalism, a problem emerges.

Since the formulation (A) is used in the imaginary time formalism, the formulation (A) with real time agrees with a result from the imaginary time formalism straightforwardly. On the other hand, the formulation (B) with real time does not agree with a result from the imaginary time formalism directly. Thus, most results in the formulation (A) and (B) are not corresponding.

We show topics that relate to a spectral function\cite{rf:bellac}\cite{rf:lands}. Since it is useful to use a spectral function at finite temperature and density\cite{rf:mallik}\cite{rf:wang}\cite{rf:harada}, an attention for the use might be important. We show formulations (A) and (B) in section\ \ref{sec:1}. In section\ \ref{sec:2}, we explain the cause of the discrepancy intuitively. In section\ \ref{sec:3}, we perform a specific calculation using the formulation (A) and (B), and we show the difference on a calculational procedure.

\vspace{1em}
\vspace{1em}
\vspace{1em}
\vspace{1em}
\section{Fermion propagator by a spectral function}\label{sec:1}

\subsection{Formulation (A)}

The ensemble average of an operator at temperature $ T=1/\beta$ and chemical potential $\mu$ is defined as

\vspace{1em}
\begin{equation}
\langle\hat{O}\rangle_{\beta}=Z^{-1}\mathrm{tr}[e^{-\beta(\hat{H}-\mu\hat{N})}\hat{O}], Z=\mathrm{tr}[e^{-\beta(\hat{H}-\mu\hat{N})}],
\end{equation}

\vspace{1em}
\noindent where $\hat{H}$ and $\hat{N}$ are a Hamiltonian and a number operator. Since we use a fermion field and the time contour $-\infty\sim+\infty$ (in our purpose it is sufficient to consider only this time contour), the time ordered propagator is defined as

\vspace{1em}
\begin{align}
S(x,y)=&\theta(x_{0}-y_{0})\langle\hat{\psi}(x)\hat{\overline{\psi}}(y)\rangle_{\beta}-\theta(y_{0}-x_{0})\langle\hat{\overline{\psi}}(y)\hat{\psi}(x)\rangle_{\beta}\nonumber\\[0.21cm]
\vspace{1em}
=&\theta(x_{0}-y_{0})S^{>}(x,y)+\theta(y_{0}-x_{0})S^{<}(x,y),\label{eq:propa1}
\end{align}

\vspace{1em}
\noindent The advanced and the retarded green functions are defined as

\begin{subequations}

\begin{align}
&S_{A}(x,y)=i\theta(x_{0}-y_{0})(S^{>}(x,y)-S^{<}(x,y)),\label{eq:adva1}\\[0.21cm]
\vspace{1em}
&S_{R}(x,y)=-i\theta(y_{0}-x_{0})(S^{>}(x,y)-S^{<}(x,y)).\label{eq:retad1}
\end{align}

\noindent In addition, we define the thermal green function as

\vspace{1em}
\begin{equation}
S_{\beta}(\tau,\bm{x},\tau^{\prime},\bm{y})=\theta(\tau-\tau^{\prime})\langle\hat{\psi}(\tau,\bm{x})\hat{\overline{\psi}}(\tau^{\prime},\bm{y})\rangle_{\beta}-\theta(\tau^{\prime}-\tau)\langle\hat{\overline{\psi}}(\tau^{\prime},\bm{y})\hat{\psi}(\tau,\bm{x})\rangle_{\beta}\label{eq:therm}
\end{equation}

\end{subequations}

\vspace{1em}
\noindent where $\tau$ is imaginary time $(\tau=it, 0\leq\tau\leq\beta)$.

We consider a case that the time evolution is given by

\vspace{1em}
\[
\hat{\psi}(t,\bm{x})=e^{i(\hat{H}-\mu\hat{N})t}\hat{\psi}(t=0,\bm{x})e^{-i(\hat{H}-\mu\hat{N})t}.
\]

\vspace{1em}
\noindent By this definition, the Kubo-Martin-Schwinger (KMS) condition for a c-number fermion field is

\vspace{1em}
\begin{equation}
\psi(t)=-\psi(t-i\beta).\label{eq:KMS1}
\end{equation}

\vspace{1em}
\noindent The KMS condition for $S^{>}(t,\bm{x})$ is

\vspace{1em}
\begin{equation}
S^{>}(t,\bm{x})=-S^{<}(t+i\beta,\bm{x}),
\end{equation}

\vspace{1em}
\noindent and $S^{>}(p)$ has the relation,

\vspace{1em}
\begin{equation}
S^{>}(p)=\displaystyle \int_{-\infty}^{\infty}dt\int d^{3}xe^{ipx}S^{>}(t,\bm{x})=-e^{\beta p_{0}}S^{<}(p).\label{eq:1}
\end{equation}

\vspace{1em}
We define the spectral function in the momentum representation as\cite{rf:bellac}\cite{rf:lands}

\vspace{1em}
\begin{equation}
\rho(p)=S^{>}(p)-S^{<}(p).
\end{equation}

\vspace{1em}
\noindent Using this spectral function and (\ref{eq:1}),

\vspace{1em}
\begin{equation}
S^{>}(p)=(1-n_{F}(p_{0}))\rho(p),\ S^{<}(p_{0})=-n_{F}(p_{0})\rho(p),\label{eq:nf1}
\end{equation}

\vspace{1em}
\noindent with

\vspace{1em}
\[
n_{F}(p_{0})=\frac{1}{e^{\beta p_{0}}+1}.
\]

\vspace{1em}
\noindent The advanced, the retarded and the thermal green functions can be expressed by a spectral function. An explicit expression for spectral function is given by\cite{rf:fetter}

\vspace{1em}
\begin{equation}
\displaystyle \rho(p_{0})=2\pi Z^{-1}\sum_{n,m}e^{-\beta K_{n}}(1+e^{\beta(K_{n}-K_{m})})\langle m|\overline{\psi}(0)|n\rangle\langle n|\psi(0)|m\rangle\delta(p_{0}+K_{n}-K_{m}),
\end{equation}

\vspace{1em}
\noindent where $K_{n}=E_{n}-\mu N_{n}.\ E_{n}$ and $N_{n}$ are eigenvalues of the operator $\hat{p}_{0}$ and $\hat{N}_{n}$ for eigenstates $|n\rangle$. We ignore spatial components for simplicity. Using this expression, the advanced, the retarded, and the thermal green functions with the spectral function are written by

\begin{subequations}

\begin{align}
&S_{A}(p_{0},\displaystyle \bm{p})=-\frac{1}{2\pi}\int_{-\infty}^{\infty}dz\frac{\rho(z,\bm{p})}{p_{0}-z+i\eta},\label{eq:As1}\\[0.21cm]
\vspace{1em}
&S_{R}(p_{0},\displaystyle \bm{p})=-\frac{1}{2\pi}\int_{-\infty}^{\infty}dz\frac{\rho(z,\bm{p})}{p_{0}-z-i\eta},\label{eq:Rs1}\\[0.21cm]
\vspace{1em}
&S_{\beta}(i\displaystyle \omega_{n},\bm{p})=-\frac{1}{2\pi}\int_{-\infty}^{\infty}dz\frac{\rho(z,\bm{p})}{p_{0}-z}.\label{eq:Ts1}
\end{align}

\end{subequations}

\vspace{1em}
The free spectral function is deriven from the free Dirac equation. Since the Hamiltonian at a zero chemical potential is replaced by $H-\mu N$, the free Dirac equation at non-zero chemical potential is written by

\vspace{1em}
\begin{equation}
(i\gamma_{0}\partial_{0}+\gamma_{0}\mu+i\bm{\gamma}\cdot\nabla-m)\psi(x)=0.\label{eq:dirac1}
\end{equation}

\vspace{1em}
\noindent Thus, (\ref{eq:propa1}), (\ref{eq:adva1}), and (\ref{eq:retad1}) for the free theory are constructed by $\psi(x)$ satisfying (\ref{eq:KMS1}) and (\ref{eq:dirac1}). $S_{(0)}^{>}(t,\bm{x})$ obeys the equation,

\vspace{1em}
\begin{equation}
(i\gamma_{0}\partial_{0}+\gamma_{0}\mu+i\bm{\gamma}\cdot\nabla-m)S_{(0)}^{>}(t,\bm{x})=0,\label{eq:S1}
\end{equation}

\vspace{1em}
\noindent Taking account of (\ref{eq:dirac1}), we can obtain the spectral function for the free theory,

\vspace{1em}
\begin{equation}
\rho^{(0)}(p)=2\pi((p_{0}+\mu)\gamma_{0}-\bm{\gamma}\cdot \bm{p}+m)\epsilon(p_{0}+\mu)\delta((p_{0}+\mu)^{2}-|\bm{p}|^{2}-m^{2}).\label{eq:spec1}
\end{equation}

\vspace{1em}
\noindent where $\epsilon(p_{0})=\theta(p_{0})-\theta(-p_{0})$.

\vspace{1em}
\vspace{1em}
\subsection{Formulation (B)}

The definitions of the green functions are the same as (\ref{eq:propa1}), (\ref{eq:adva1}), (\ref{eq:retad1}), and (\ref{eq:therm}). However, a fermion field is not same. We consider the case that the time evolution is given by

\vspace{1em}
\begin{equation}
\hat{\psi}^{\prime}(t,\bm{x})=e^{i\hat{H}t}\hat{\psi}^{\prime}(t=0,\bm{x})e^{-i\hat{H}t}.
\end{equation}

\vspace{1em}
\noindent Then, the KMS condition is given by

\vspace{1em}
\begin{equation}
\psi^{\prime}(t,\bm{x})=-e^{\beta\mu}\psi^{\prime}(t-i\beta,\bm{x}),\label{eq:KMS2}
\end{equation}

\vspace{1em}
\noindent and

\vspace{1em}
\begin{equation}
S^{>\prime}(t,\bm{x})=-e^{\beta\mu}S^{<\prime}(t+i\beta,\bm{x}).
\end{equation}

\vspace{1em}
\noindent Then, $S^{>\prime}(p)$ has the relation,

\vspace{1em}
\begin{equation}
S^{>\prime}(p)=\displaystyle \int_{-\infty}^{\infty}dt\int d^{3}xe^{ipx}S^{>\prime}(t,\bm{x})=-e^{\beta(p_{0}-\mu)}S^{<\prime}(p).\label{eq:2}
\end{equation}

\vspace{1em}
\noindent Since the definition for a spectral function is same, $S^{>\prime}(p)$ and $S^{<\prime}(p)$ can be expressed by

\vspace{1em}
\begin{equation}
S^{>\prime}(p)=(1-n_{F}(p_{0}-\mu))\rho(p),\ S^{<\prime}(p)=-n_{F}(p_{0}-\mu)\rho(p),\label{eq:nf2}
\end{equation}

\vspace{1em}
\noindent with

\vspace{1em}
\[
n_{F}(p_{0}-\mu)=\frac{1}{e^{\beta(p_{0}-\mu)}+1}.
\]

\vspace{1em}
An explicit expression for the spectral function is given by\cite{rf:mallik}

\vspace{1em}
\begin{equation}
\displaystyle \rho^{\prime}(p_{0})=2\pi Z^{-1}\sum_{n,m}e^{-\beta(E_{n}-\mu N_{n})}(1+e^{\beta(E_{n}-E_{m}+\mu)})\langle m|\overline{\psi}(0)|n\rangle\langle n|\psi(0)|m\rangle\delta(p_{0}+E_{n}-E_{m}).
\end{equation}

\vspace{1em}
\noindent If we perform the Fourier transform,

\vspace{1em}
\begin{equation}
f(p_{0})=\displaystyle \int_{-\infty}^{\infty}dte^{ip_{0}t}f(t),\label{eq:four1}
\end{equation}

\vspace{1em}
\noindent the retarded green function is

\vspace{1em}
\begin{equation}
S_{R}^{\prime}(p_{0},\displaystyle \bm{p})=-\frac{1}{2\pi}\int_{-\infty}^{\infty}dz\frac{\rho^{\prime}(z,\bm{p})}{p_{0}-z+i\eta}.\label{eq:Rs2}
\end{equation}

\vspace{1em}
\noindent As can be expected from (\ref{eq:nf2}) and (\ref{eq:spec2}), this does not agree with (\ref{eq:Rs1}). On the other hand, using the different Fourier transform,

\vspace{1em}
\begin{equation}
f(p_{0})=\displaystyle \int_{-\infty}^{\infty}dte^{ip_{0}t}e^{i\mu t}f(t),\label{eq:four2}
\end{equation}

\vspace{1em}
\noindent one obtains the result,

\vspace{1em}
\begin{equation}
S_{R}(p_{0},\displaystyle \bm{p})=-\frac{1}{2\pi}\int_{-\infty}^{\infty}dz\frac{\rho^{\prime}(z,\bm{p})}{p_{0}+\mu-z+i\eta}.\label{eq:Rs3}
\end{equation}

\vspace{1em}
\noindent If $\rho(z,\bm{p})$ depends on $ z+\mu$, (\ref{eq:Rs3}) relates to (\ref{eq:Rs1}) by the change of variable (see also (\ref{eq:nf1}) and (\ref{eq:nf2})). Thus, using not (\ref{eq:four1}) but (\ref{eq:four2}), an equivalent result is derived. This replacement is equal to embedding (\ref{eq:rel}) in the Fourier transform. 

The free Dirac equation at non-zero chemical potential is

\vspace{1em}
\begin{equation}
(i\gamma_{0}\partial_{0}+i\bm{\gamma}\cdot\nabla-m)\psi^{\prime}(x)=0.\label{eq:dirac2}
\end{equation}

\vspace{1em}
\noindent The Dirac equation in the formulation (B) is the same as the zero chemical potential form. From (\ref{eq:KMS1}), (\ref{eq:dirac1}), (\ref{eq:KMS2}) and (\ref{eq:dirac2}), $\psi(x)$ and $\psi^{\prime}(x)$ have the relation,

\vspace{1em}
\begin{equation}
\psi(x)=e^{i\mu t}\psi^{\prime}(x).\label{eq:rel}
\end{equation}

\vspace{1em}
\noindent$S_{(0)}^{>\prime}(t,\bm{x})$ constructed by $\psi^{\prime}(x)$ must satisfy the equation,

\vspace{1em}
\begin{equation}
(i\gamma_{0}\partial_{0}+\bm{\gamma}\cdot\nabla-m)S_{(0)}^{>\prime}(t,\bm{x})=0.\label{eq:S2}
\end{equation}

\vspace{1em}
\noindent Taking account of (\ref{eq:dirac2}), we can obtain the spectral function for the free theory,

\vspace{1em}
\begin{equation}
\rho^{(0)\prime}(p)=2\pi(p_{0}\gamma_{0}-\bm{\gamma}\cdot \bm{p}+m)\epsilon(p_{0})\delta(p_{0}^{2}-\bm{p}^{2}-m^{2}).\label{eq:spec2}
\end{equation}

\vspace{1em}
\noindent As pointed out above, this result is consistent with (\ref{eq:spec1}).

\vspace{1em}
\vspace{1em}
\section{Cause of the discrepancy}\label{sec:2}

Obtaining a different result is understood from (\ref{eq:rel}). More specifically, the different results from a plane wave solution. Plane wave solutions for (\ref{eq:dirac1}) and (\ref{eq:dirac2}) are

\vspace{1em}
\begin{equation}
\psi(t)\sim e^{-i(p_{0}-\mu)t},\label{eq:plane1}
\end{equation}

\vspace{1em}
\noindent and

\vspace{1em}
\begin{equation}
\psi^{\prime}(t)\sim e^{-ip_{0}t},\label{eq:plane2}
\end{equation}

\vspace{1em}
\noindent respectively. Thus, there is a difference of $\mu$ for a frequency (energy) between (\ref{eq:dirac1}) and (\ref{eq:dirac2}). Owing to this, after performing the same Fourier transform, the formulation (A) and (B) derive a different result.

Using (\ref{eq:plane1}), $S^{>}(x)$ in (\ref{eq:S1}) is written as

\vspace{1em}
\begin{equation}
S^{>}(x)=\displaystyle \int\frac{d^{3}p}{(2\pi)^{3}}\frac{1}{2E_{\bm{p}}}(i\gamma_{0}(\partial_{0}-i\mu)+i\bm{\gamma}\cdot\nabla+m)e^{i\mu t}[(1-n_{F}(E_{\bm{p}}-\mu))e^{-ipx}-n_{F}(E_{\bm{p}}+\mu)e^{ipx}],\label{eq:plane11}
\end{equation}

\vspace{1em}
\noindent where $p_{0}=E_{\bm{p}}=\sqrt{\bm{p}^{2}+m^{2}}$. On the other hand, using Eq.\ (\ref{eq:plane2}), $S^{>\prime}(x)$ in (\ref{eq:S2}) is written as

\vspace{1em}
\begin{equation}
S^{>\prime}(x)=\displaystyle \int\frac{d^{3}p}{(2\bm{\pi})^{3}}\frac{1}{2E_{\bm{p}}}(i\partial\hspace{-.50em}/+m)[(1-n_{F}(E_{\bm{p}}-\mu))e^{-ipx}-n_{F}(E_{\bm{p}}+\mu)e^{ipx}].\label{eq:plane21}
\end{equation}

\vspace{1em}
\noindent These satisfy each KMS condition. Since there is an extra factor $e^{i\mu t}$ in (\ref{eq:plane11}), a different result is derived on a calculation of a Feynman diagram if one calculates straightforwardly. However, the difference is removed easily in the momentum representation using the Fourier transform (\ref{eq:four2}).

\vspace{1em}
\vspace{1em}
\section{Example}\label{sec:3}

\subsection{Calculation of the summation by (A)}

We show a procedure of a sum over the Matsubara frequency by using a spectral function\cite{rf:bellac}. A sum over the Matsubara frequency can be done easily by this procedure. We calculate the summation,

\vspace{1em}
\begin{equation}
I=T\displaystyle \sum_{m}\omega_{m}G_{\beta}(\omega_{m},E_{1})G_{\beta}(\omega_{n}-\omega_{m},E_{2}),\label{eq:sum1}
\end{equation}

\vspace{1em}
\noindent where $\omega_{n}$ and $\omega_{m}$ are the boson and the fermion Matsubara frequency, respectively. For simplicity, we use

\vspace{1em}
\begin{equation}
G_{\beta}(\displaystyle \omega_{m},E_{1})=\frac{1}{-(i\omega_{m}+\mu)^{2}+E_{1}^{2}},\ G_{\beta}(\omega_{n}-\omega_{m},E_{2})=\frac{1}{-(i\omega_{n}-(i\omega_{m}+\mu))^{2}+E_{2}^{2}}.
\end{equation}

\vspace{1em}
\noindent$(p_{0}+\mu)\gamma_{0}-\bm{\gamma}\cdot \bm{p}+m\ \ $in (\ref{eq:spec1}) is removed. The relation,

\vspace{1em}
\begin{equation}
i\displaystyle \int_{0}^{\beta}d\tau\frac{dG(\tau,\bm{p})}{d\tau}e^{i\omega_{m}\tau}=\omega_{m}\int_{0}^{\beta}d\tau G(\tau,\bm{p})e^{i\omega_{m}\tau},\label{eq:sum2}
\end{equation}

\vspace{1em}
\noindent can be derived from

\vspace{1em}
\begin{equation}
G_{\beta}(i\displaystyle \omega_{m},\bm{p})=\int_{0}^{\beta}d\tau G(\tau,\bm{p})e^{i\omega_{m}\tau},
\end{equation}

\vspace{1em}
\noindent and (\ref{eq:KMS1}) changed by imaginary time $\tau$. Substituting (\ref{eq:sum2}) into $I$, and using (\ref{eq:nf1}),

\vspace{1em}
\begin{align}
I=&\ iT\displaystyle \sum_{m}\int_{0}^{\beta}d\tau d\tau^{\prime}\frac{dG(\tau,E_{1})}{d\tau}G(\tau^{\prime},E_{2})e^{i\omega_{m}\tau}e^{i(\omega_{n}-\omega_{m})\tau^{\prime}}\nonumber\\[0.21cm]
\vspace{1em}
=&\ i\displaystyle \int_{-\infty}^{\infty}\frac{dp_{0}}{2\pi}\int_{-\infty}^{\infty}\frac{dk_{0}}{2\pi}k_{0}\frac{1-n_{F}(p_{0})-n_{F}(k_{0})}{i\omega_{n}-p_{0}-k_{0}}\sigma_{+}^{(0)}(k_{0},E_{1})\sigma_{-}^{(0)}(p_{0},E_{2})\label{eq:sum3}
\end{align}

\vspace{1em}
\noindent where

\vspace{1em}
\[
\sigma_{+}^{(0)}(k_{0},E_{1})=2\pi\epsilon(k_{0}+\mu)\delta((k_{0}+\mu)^{2}-E_{1}^{2}),
\]

\vspace{1em}
\[
\sigma_{-}^{(0)}(p_{0},E_{2})=2\pi\epsilon(p_{0}-\mu)\delta((p_{0}-\mu)^{2}-E_{2}^{2})
\]

\vspace{1em}
\vspace{1em}
\subsection{Calculation of the summation by (B)}

We consider the same summation (\ref{eq:sum1}). However, the same Fourier transform and a spectral function are not available. Using the Fourier transform (\ref{eq:four2}), (\ref{eq:sum2}) turns into

\vspace{1em}
\begin{equation}
\displaystyle \omega_{m}\int_{0}^{\beta}d\tau S(\tau,E_{1})e^{i(\omega_{m}-i\mu)\tau}=i\int_{0}^{\beta}d\tau\frac{dS(\tau,E_{1})}{d\tau}e^{i(\omega_{m}-i\mu)\tau}+i\mu\int_{0}^{\beta}d\tau S(\tau,E_{1})e^{i(\omega_{m}-i\mu)\tau}.
\end{equation}

\vspace{1em}
\noindent Substituting this into (\ref{eq:sum1}),

\vspace{1em}
\begin{align*}
I=&\ iT\displaystyle \sum_{m}\int_{0}^{\beta}d\tau d\tau^{\prime}\frac{dS(\tau,E_{1})}{d\tau}S(\tau^{\prime},E_{2})e^{(i\omega_{m}+\mu)\tau}e^{(i\omega_{n}-i\omega_{m}-\mu)\tau^{\prime}}\nonumber\\[0.21cm]
\vspace{1em}
&+i\displaystyle \mu T\sum_{m}\int_{0}^{\beta}d\tau d\tau^{\prime}S(\tau,E_{1})S(\tau^{\prime},E_{2})e^{(i\omega_{m}+\mu)\tau}e^{(i\omega_{n}-i\omega_{m}-\mu)\tau^{\prime}}\nonumber\\[0.21cm]
\vspace{1em}
=&\ i\displaystyle \int_{-\infty}^{\infty}\frac{dk_{0}}{2\pi}\int_{-\infty}^{\infty}\frac{dp_{0}}{2\pi}(k_{0}-\mu)\frac{1-n_{F}(k_{0}-\mu)-n_{F}(p_{0}+\mu)}{i\omega_{n}-p_{0}-k_{0}}\sigma(k_{0},E_{1})\sigma(p_{0},E_{2}),
\end{align*}

\vspace{1em}
\noindent where

\vspace{1em}
\begin{equation}
\sigma(k_{0},E_{1})=2\pi\epsilon(k_{0})\delta(k_{0}^{2}-E_{1}^{2}).
\end{equation}

\vspace{1em}
\noindent This result agrees with (\ref{eq:sum3}). Note that (\ref{eq:nf2}) and (\ref{eq:spec2}) are used. Simultaneous usage of the formulation (A) and the modified Fourier transform produces a incorrect result.

\vspace{1em}
\subsection{Spectral function in the 1-loop order}

We consider the scalar boson-fermion interaction. We perform the summation without using the Fourier transform to reduce argument. Ignoring contributions of vertices, the 1-loop fermion self energy in the imaginary formalism is given by

\vspace{1em}
\begin{equation}
\displaystyle \Sigma(i\omega_{n},\bm{p})=-T\sum_{l}\int\frac{d^{3}k}{(2\pi)^{3}}S_{\beta}(i\omega_{l},\bm{k})D_{\beta}(i\omega_{n}-i\omega_{l},\bm{p}-\bm{k}),
\end{equation}

\vspace{1em}
\noindent where $\omega_{n}$ and $\omega_{l}$ are the fermion Matsubara frequency. After replacing $S_{\beta}$ and $D_{\beta}$ with (\ref{eq:Ts1}), performing the sum over the Matsubara frequency by a summation formula without using the Fourier transform,

\vspace{1em}
\begin{equation}
\displaystyle \Sigma(i\omega_{n},\bm{p})=-\int\frac{d^{3}k}{(2\pi)^{3}}\int_{-\infty}^{\infty}\frac{dz_{1}}{2\pi}\frac{dz_{2}}{2\pi}\rho_{F}(z_{1},\bm{k})\rho_{B}(z_{2},\bm{p}-\bm{k})\frac{1}{i\omega_{n}-z_{1}-z_{2}}\big(n_{F}(z_{1})+n_{B}(-z_{2})\big),
\end{equation}

\vspace{1em}
\noindent where $n_{B}(z)=1/(e^{\beta z}-1),\ \rho_{F}$ and $\rho_{B}$ are the fermion and the boson spectral functions, respectively. After analytic continuation $ i\omega_{n}\rightarrow p_{0}+i\eta$ to an arbitrary continuous value $p_{0}$, the imaginary part is given by

\vspace{1em}
\begin{equation}
{\rm Im}\displaystyle \Sigma_{R}(p_{0},\bm{p})=\frac{1}{4\pi}\int\frac{d^{3}k}{(2\pi)^{3}}\int_{-\infty}^{\infty}dz_{1}dz_{2}\rho_{F}(z_{1},\bm{k})\rho_{B}(z_{2},\bm{p}-\bm{k})[n_{F}(z_{1})+n_{B}(-z_{2})]\delta(p_{0}-z_{1}-z_{2})
\end{equation}

\vspace{1em}
\noindent Using free spectral functions (\ref{eq:spec2}) and $\rho_{B}^{(0)}(q)=2\pi\epsilon(q_{0})\delta(q^{2}-m_{B}^{2})$,

\vspace{1em}
\begin{align}
{\rm Im}\displaystyle \Sigma_{R}(p_{0},\bm{p})=\frac{-\pi}{4E_{1}E_{2}}&\displaystyle \int\frac{d^{3}k}{(2\pi)^{3}}(E_{1}\gamma_{0}-\bm{\gamma}\cdot \bm{k}+m)\nonumber\\[0.21cm]
\vspace{1em}
&\times\big[(1-n_{F}(E_{1}-\mu)+n_{B}(E_{2}))\delta(p_{0}+\mu-E_{1}-E_{2})\nonumber\\[0.21cm]
\vspace{1em}
&\hspace{1.5em}+(n_{F}(E_{1}-\mu)+n_{B}(E_{2}))\delta(p_{0}+\mu-E_{1}+E_{2})\nonumber\\[0.21cm]
\vspace{1em}
&\hspace{1.5em}-(1-n_{F}(E_{1}+\mu)+n_{B}(E_{2}))\delta(p_{0}+\mu+E_{1}+E_{2})\nonumber\\[0.21cm]
\vspace{1em}
&\hspace{1.5em}-(n_{F}(E_{1}+\mu)+n_{B}(E_{2}))\delta(p_{0}+\mu+E_{1}-E_{2})\big]
\end{align}

\vspace{1em}
\noindent where $E_{1}=\sqrt{|\bm{k}|^{2}+m^{2}},\ E_{2}=\sqrt{|\bm{p}-\bm{k}|^{2}+m_{B}^{2}}$. The 1-loop fermion spectral function can be obtain by\cite{rf:bellac}\cite{rf:lands}

\vspace{1em}
\begin{equation}
\rho_{F}^{(1)}(p)=-i(\Sigma(p_{0}+i\eta)-\Sigma(p-i\eta))=2{\rm Im}\Sigma_{R}(p),
\end{equation}

\vspace{1em}
\noindent The delta functions in $\rho_{F}^{(1)}(p)$ obtained strictly by the imaginary time formalism are different from (50) in Ref \cite{rf:mallik}, which used the formulation (B). Thus, the method in Ref \cite{rf:mallik} does not correspond to the imaginary time formalism. The source of mistake is the Fourier transform (23) in Ref \cite{rf:mallik}. If one wants to correspond to the imaginary time formalism, $p_{0}$ must be replaced by $ p_{0}+\mu$. However, when one does not consider corresponding to the imaginary time formalism, the method in Ref \cite{rf:mallik} is correct. (As above mentioned, it is nothing more than a shift of an energy.) In other words, the difference of $\mu$ is absorbed into an external line $p_{0}$ to regard $\Sigma_{R}(p_{0},\bm{p})$ as $\Sigma_{R}(p_{0}^{\prime}=p_{0}+\mu,\bm{p})$. In fact, using (\ref{eq:Rs3}), the same retarded green function can be obtained.

\vspace{1em}
\vspace{1em}
\section{Summary}

There are two formulations at non-zero chemical potential. One is the formulation (A) that a Lagrangian includes a chemical potential, the other is the formulation (B) that a Lagrangian does not include a chemical potential. The former corresponds to the imaginary time formalism more directly.

It is necessary to note that a calculational procedure in two formulations is different. For example, $S_{(0)}^{>}(x)$ in (A) has a factor $e^{i\mu t}$ as compared with $S_{(0)}^{>\prime}(x)$ in (B). This difference is important when performing a summation in imaginary time formalism and calculating a Feynman diagram in the real time formalism, etc. A wrong choice makes a mistake. Owing to this, it is necessary to understand the formulation used, combining (A) and (B) carelessly is unsafe. In particular, it is important to note that the advanced, the retarded, and the thermal green functions expressed by a spectral function are different.

(B) does not correspond to the imaginary time formalism directly because a shift of an energy exists. This shift is removed by modifying the Fourier transform. The modified Fourier transform is equal to embedding (\ref{eq:rel}) in the ordinary Fourier transform. Using the modified Fourier transform, a correspondence between (A) and (B) during a calculation becomes easier, (B) agrees with the imaginary time formalism.

Incidentally, it should be seen that (A) and (B) in the real time functional integral formulation are related by the canonical transform\cite{rf:weldon2}. Treating (B) as the canonical transform of (A), a source term has the factor $e^{i\mu t}$. Thus, the time ordered propagator obtained by a functional derivative in (B) corresponds to the one in (A).

\vspace{1em}
\end{document}